\documentstyle[prl,twocolumn,aps,epsf]{revtex}

\begin{document}

\title{Boundary interactions changing operators and dynamical
correlations
in quantum impurity problems.}
\author{F. Lesage, H. Saleur$^*$.}
\address{Department of Physics, University of Southern California,
Los-Angeles, CA 90089-0484.}
\date{\today}
\maketitle

\begin{abstract}
Recent developments have made  possible the computation of
equilibrium
dynamical correlators in quantum impurity problems.
In many situations however,
one is rather interested in  correlators subject to
a non equilibrium initial preparation; this is the case for
instance for the occupation probability $P(t)$ in  the double well
problem of dissipative quantum mechanics (DQM). We show in this paper
 how to handle
this situation
 in the framework of integrable quantum field theories
by introducing ``boundary interactions changing operators''.
We determine the properties of these operators by using an axiomatic
approach similar
in spirit to what is done for form-factors. This allows us to obtain
new exact
results for $P(t)$; for instance, we find that
 that at large times (or small $g$), the leading behaviour  for
$g<{1\over 2}$
is
$P(t)\propto e^{-\Gamma t}\cos\Omega t$, with the universal ratio.
$\frac{\Omega}{\Gamma}=\cot\frac{\pi g}{2(1-g)}$.

\end{abstract}

\smallskip

Strongly correlated electron  systems, which have been the subject of
intense
study in recent years, pose many theoretical challenges
due to their  essentially non perturbative nature. The greatest
progress have
been realized for  systems that are one dimensional, in particular
quantum
impurity problems. Fixed points and their vicinity can  then be
investigated
with techniques of conformal field theory (CFT) \cite{ian}, and the
behaviour
between fixed points with the Bethe ansatz
\cite{bethe},\cite{fendley}. Even with these powerful tools,
the understanding of dynamical properties is still quite incomplete.
For instance
the Bethe ansatz provides thermodynamic properties, and, with some
effort, DC properties, including   out of equilibrium
\cite{fendleyi}; but
time and space dependent correlations, though accessible in principle
\cite{korepin}, have remained largely undetermined until recently. It
is
highly desirable to make progress in that area, in particular in view
of
recent  exciting experiments \cite{exps},\cite{expsi}

In the last two years, building on formal works on integrable massive
quantum
field theories (QFTs) \cite{ghozamo},{\cite{smirnov},\cite{giuseppe},
it has become possible
to determine  some  of these correlators exactly (ie with
arbitrary accuracy, all the way from short to large distances)
\cite{skorik}.
The results obtained sometimes exhibit striking behaviour,
illustrating
the difficulty to build an intuition for non perturbative systems.
For instance,
in the double well problem of dissipative quantum mechanics, it was
found that the transition from coherent to incoherent regime,
if based on the two spin correlation function,
takes place at a value of the dissipation $\alpha={1\over 3}$
\cite{skorik},\cite{kosti}, in  contrast with
the value $\alpha={1\over 2}$ expected before
\cite{sudip},\cite{sudipi}.  This result later raised questions about
which quantity should be used to describe this transition
\cite{eggr}.

Unfortunately, correlators that can be computed in the framework of
QFT are not always the physically relevant ones. This is
because natural objects in QFT are vacuum vacuum Green functions,
while in many experimental settings, what is measurable are time
evolutions
for a system prepared in a given state that is, in fact, not even an
eigenstate.
A well known example of this is the occupation probability $P(t)$ in
the double
 well
problem of DQM \cite{sudip}. In that problem,
the system has a non equilibrium initial preparation:
for
negative times, the spin is held fixed in the state $S_z=1$, with an
equilibrated
environment. At time $t=0$, the constraint is released, and the
dynamics starts
from $P(0)=1$ with a factorized spin-environment initial state. The
behaviour
 of
$P(t)$ has been the subject of much work lately
\cite{Egger},\cite{Strong}.
Several possibilities for measuring this quantity
 experimentally have also been proposed
\cite{seethna},\cite{leclair}. On the other
hand, no exact result for $P(t)$ has been available up to now, except
at $\alpha={1\over 2}$
\cite{sudip}.

In this letter, we introduce a  method to compute correlators in
prepared
initial
states that are not eigenstates. The key ingredient is a
generalization
of the  boundary conditions changing operators of CFTs
\cite{ian},\cite{JCardy}
to theories with boundary interactions. We illustrate the method by
the computation of
$P(t)$. The final results are too bulky to be presented here;
they have the following  interesting features. For $g>{1\over 2}$,
the behaviour of $P( t)$
is incoherent, and $P(t)$ has the general form
\begin{equation}
P(t)=\sum_{n=1}^\infty a_n e^{-2n T_b t}\label{poft}
\end{equation}
where $\sum a_n=1$, and $T_b$ is an energy scale characterizing the
tunnel splitting of
the free system (see below). The transition from coherent to
incoherent regime
takes place at $g={1\over 2}$, and is  related with the transition
from repulsive to attractive regime in the sine-Gordon model. For
$g<{1\over 2}$,
$P(t)$ has a form similar to (\ref{poft}) but the sum involves
also (known) complex arguments in the exponential. At large times,
the dominant behaviour is
\begin{equation}
P(t)\simeq \exp\left[ -2 t T_b \sin^2\frac{\pi g}{2(1-g)}
\right] \cos\left[ t T_b \sin\frac{\pi g}{1-g}\right]
\label{poftasymp}
\end{equation}
which in turn leads to the prediction for the ratio of the damping
factor to the period of oscillations
\begin{equation}
\frac{\Omega}{\Gamma}=\cot\frac{\pi g}{2(1-g)}.
\label{ratio}
\end{equation}
All these results are new, except for the fact that the transition
occurs right at
$g={1\over 2}$, which was demonstrated using perturbation
 around $g={1\over 2}$ in \cite{Egger}.

To start, we  recall the concept of boundary conditions
changing operator,
and we introduce their generalization, in the simple case of the
Ising model (closely related to the double well problem at $g={1\over
2}$). We consider the massive  Ising model with an inhomogeneous
boundary magnetic field
\begin{eqnarray}
\label{action}
A&=&\int_{-\infty}^0 dx \int_{-\infty}^\infty dy
a_{FF}(x,y) \\ &+&{1\over 2}
\int_{-\infty}^\infty
dy\left[\left(\psi\bar{\psi}\right)(x=0)+a\dot{a}\right]
+\int_{-\infty}^\infty dyh(y)\sigma_B(y),
\nonumber
\end{eqnarray}
where $a_{FF}$ is  the usual massive free Majorana fermion action,
$a$ is a boundary fermion satisfying $a^2=1$, $\sigma_B$
is the boundary spin operator, which coincides with
${1\over 2}\left(\psi+\bar{\psi}\right)a$.
If the mass vanishes, the theory is conformal invariant in the bulk.
Conformal invariant boundary conditions
are then  obtained with $h=0$ (free spins)
or $h=\pm\infty$ (fixed spins). As shown by Cardy \cite{JCardy}, the
situation
where  boundary conditions are conformal invariant
but inhomogeneous can be represented by the insertion of
conformal operators on the boundary. For instance, the case with
$h=0$ for $y\leq 0$
and $h=\infty $ for $y>0$ is described by  the insertion of the
operator $\Phi_{12}$ of
dimension $\Delta={1\over 16}$ at $y=0$.

Let us now consider a situation where conformal invariance is broken:
first suppose  that the bulk mass is non zero, and that
the boundary magnetic field is uniform and  equal to $h_a$. This can
be handled
 using the integrability of the problem \cite{Barry},\cite{ghozamo}.  
The
hamiltonian
(for time evolution along the $y$ direction) is diagonalized using a
basis
of multiparticle states (here, simply fermions) that have factorized
scattering
both in the bulk and at the boundary. We use in the following
rapidity variables
parametrizing energy and momentum as $e=m\cosh\theta,
p=m\sinh\theta$; the
asymptotic  states have  then the general form
$|\theta_1,\ldots,\theta_n\rangle_a$. $S$ and $R$  matrix elements
relate these
states with others, where some of the rapidities have been switched,
or had their
sign changed. Using the known $S$ and $R$ matrices \cite{ghozamo},
correlators
can then be computed \cite{skorik},\cite{giuseppei}, provided one
knows the matrix elements
of the operators (form-factors) in the multiparticle basis
\cite{smirnov}. For instance,
for the energy density in the Ising model, the only non vanishing
form-factors
are
\begin{equation}
\langle \epsilon(w,\bar{w})\rangle
={m\over 2}\int {d\theta\over 2\pi}
R_a(\theta)\cosh\theta e^{-2imx\sinh\theta},
\label{opte}
\end{equation}
and
\begin{eqnarray}
&{}_a&\langle 0|\epsilon(w,\bar{w})|\theta_1\theta_2\rangle_a
=im e^{-my(\cosh\theta_1+\cosh\theta_2)} \\ \nonumber
&\times & \prod_{k=1,2}
[1+R_b(\theta_k)t_k] \sinh(\frac{\theta_1-\theta_2}{2})
e^{imx(\sinh\theta_1+\sinh\theta_2)}
\label{tpte}
\end{eqnarray}
where $t_i$ acts on functions of many variables
by changing the sign of $\theta_i$, and $R_b(\theta)=
i\tanh\left({i\pi\over 4}-{\theta\over
2}\right){\kappa_a-i\sinh\theta\over
\kappa_a+i\sinh\theta}$, $\kappa_a=1-{h_a^2\over 2m}$.

The case where the theory
is still massless in the bulk is the most interesting for
applications \cite{skorik};
it can simply be obtained as a massless limit of the previous
description,
with $m\to 0$ and $\theta\to\pm\infty$.

Now suppose that the bulk mass is non zero, and in addition the
boundary
magnetic field is inhomogeneous: $h=h_a$ for $y\leq 0$ and $h=h_b$
for
$y>0$.We now have two different
hamiltonians to diagonalize, giving rise to two sorts of
multiparticle
eigenstates. Since these multiparticle eigenstates provide complete
sets of states, the scalar products
\begin{equation}
{}_b\langle \theta_n,\ldots,\theta_1
|\theta_{n+1}\ldots\theta_{n+m}\rangle_a
\label{scalar}
\end{equation}
have to be non zero in general, even for disjoint sets of
rapidities: they are nothing but the matrix elements of ``boundary
interactions
changing operators''.

Knowledge of the scalar products (\ref{scalar}) is  what is required
to compute
a quantity like $P(t)$, as we will demonstrate later. For the moment,
let us discuss
how to determine these scalar products. Consider the simplest case,
\begin{equation}
{\ _b\langle\theta_2\theta_1|0\rangle_a\over
\ _b\langle 0|0\rangle_a}=G(\theta_1,\theta_2).
\end{equation}
Since  for the Ising model, the integrable particles are just
fermions
with
$S=-1$, $G$ must satisfy the exchange  relations
\begin{eqnarray}
G(\theta_1,\theta_2)&=&-G(\theta_2,\theta_1)
=R^*_b(\theta_1)
G(-\theta_1,\theta_2) \nonumber \\
&=&R^*_b(\theta_2)G(\theta_1,-\theta_2).
\end{eqnarray}
In addition, $G$ must have a kinematical pole at
$\theta_2=\theta_1-i\pi$. To see this,
observe that there are two possible expressions for the one point
function
of the energy: for $y>0$ one has
\begin{eqnarray}
\label{epsil}
\langle\epsilon(w,\bar{w})\rangle_{ba}&=&
\nonumber
\int_0^\infty {d\theta_1d\theta_2\over
8\pi^2}\{  {}_b\langle 0|\epsilon(w,\bar{w})|
\theta_1\theta_2\rangle_b\
_b\langle\theta_2\theta_1|0\rangle_a \\
&+&\ _b\langle 0|\epsilon(w,\bar{w})|0\rangle_b\
_b\langle 0|0\rangle_a \}/_b\langle 0|0\rangle_a,
\end{eqnarray}
while for $y<0$, one has
\begin{eqnarray}
\label{epsil2}
\langle\epsilon(w,\bar{w})\rangle_{ba}&=&
\nonumber
\int_0^\infty {d\theta_1d\theta_2\over
8\pi^2}\{  _b\langle 0|\theta_1\theta_2\rangle_a \
{}_a\langle \theta_2\theta_2
|\epsilon(w,\bar{w})|0\rangle_a
\\
&+&\ _b\langle 0|0\rangle_a \
_a\langle 0|\epsilon(w,\bar{w})|0\rangle_a
 \}/_b\langle 0|0\rangle_a.
\end{eqnarray}

Replace now the form factors of $\epsilon$ by their
explicit forms:
 for (\ref{epsil2}) to be the analytical continuation
of (\ref{epsil}), we need the  residue
condition
\begin{equation}
\hbox{Res } G(\theta,\theta-i\pi)= -i \left(1
-{R_a(\theta)\over
R_b(\theta)}\right)
\end{equation}
This is enough to determine the function $G$.
 We find, using the parametrization
$\kappa_{a/b}=-\cosh\theta_{a/b}$,
\begin{eqnarray}
G&(&\theta_1,\theta_2)={i\over
8}\prod_{i=1,2}f(\theta_i)\ \sinh\theta_i\
{\kappa_b+i\sinh\theta_i\over
\kappa_b-i\sinh\theta_i}\nonumber \\
&\times&{1\over \cosh({\theta_i\over 2}+i{\pi\over 4})}
\tanh\left({\theta_1-\theta_2\over 2}\right)
\tanh\left({\theta_1+\theta_2\over 2}\right),
\end{eqnarray}
where
\begin{equation}
f(\theta)=\sqrt{-i(\kappa_a-\kappa_b)}
\Phi(\theta|\theta_b,\theta_a) \Phi(\theta|-\theta_b,-\theta_a)
\end{equation}
and
\begin{eqnarray}
\nonumber
&\Phi(\theta|\theta_{b},\theta_a)\equiv {1\over
\cosh(
{\theta-\theta_{b}\over 2}-{i\pi\over 4})}
\prod_{n=0}^\infty
{\Gamma \left({5\over 4}+n-i{\theta-\theta_{b}\over 2\pi}\right)\over
\Gamma \left({3\over 4}+n-i{\theta-\theta_{b}\over 2\pi}\right)}\\
&{\Gamma \left({5\over 4}+n+i{\theta-\theta_{b}\over
2\pi}\right)\over
\Gamma \left({3\over 4}+n+i{\theta-\theta_{b}\over 2\pi}\right)}
{\Gamma \left({3\over 4}+n-i{\theta-\theta_a\over 2\pi}\right)\over
\Gamma \left({5\over 4}+n-i{\theta-\theta_a\over 2\pi}\right)}
{\Gamma \left({3\over 4}+n+i{\theta-\theta_a\over 2\pi}\right)\over
\Gamma \left({5\over 4}+n+i{\theta-\theta_a\over 2\pi}\right)}.
\end{eqnarray}
Inspection shows that $G$ has the same poles as those of $R_b^*$.
Crossing in this problem
reads ${}_b\langle 0|\theta_1,\theta_2\rangle_a={}_b\langle
\theta_1-i\pi,
\theta_2-i\pi|0\rangle_a$. There is also a non trivial monodromy:
$G(\theta_1+2i\pi,\theta_2)=-R_a(\theta_1)
R_b^*(\theta_1)G(\theta_1,\theta_2)$.

Some comments are in order:  the general solution for
an arbitrary number of particles is not difficult to find
and will be given in\cite{nous}.
The previous expression depend only on $h_{a/b}^2$, and in fact, are
only
valid for
$h_a h_b>0$. The case of fields with opposite signs requires
an additional particle at imaginary rapidity, see \cite{nous}.

It is interesting now to get back to the conformal case: let $m\to 0$
and,
for instance,
 $h_a=\infty$
and $h_b=0$.  Setting $e=\pm p=e^{\beta}$, we
find from the foregoing formulas
\begin{equation}
{_+\langle\beta_2\beta_1|0\rangle_F
\over _+\langle0|0\rangle_F}= i\tanh{\beta_1-\beta_2\over 2},
\end{equation}
where $+$ designates up spins, $F$ free spins. This is nothing
but the form-factor of the $\Delta={1\over 16}$ operator,
in agreement with \cite{JCardy}.
We are then led  to the following spectral representation
for the energy one point function
\begin{eqnarray}
&\langle&\epsilon(w,\bar{w})\rangle_{+F}=-{1\over 4\pi x}+
\\
&i& \int
{d\beta_1d\beta_2\over
4\pi^2} \nonumber
e^{(\beta_1+\beta_2)/2} \tanh(\frac{\beta_{12}}{2})
e^{ix(e^{\beta_1}-
e^{\beta_2})}
e^{-y(e^{\beta_1}+e^{\beta_2})}.
\end{eqnarray}
Explicit evaluation gives
\begin{equation}
\langle\epsilon(w,\bar{w})\rangle_{+F}={1\over 4\pi}
\left({1\over x}
-{y\over x\sqrt{x^2+y^2}}\right)-{1\over 4\pi x}
\end{equation}
in agreement with  the result obtained in \cite{Burk} in
the context of conformal field theory.

The scalar products (\ref{scalar}) can be determined using the same
ideas
for more general integrable interacting theories.
The expressions become quite bulky and won't be reproduced here; some
very simple features determine most of $P(t)$ properties, as we now
explain.

The double well
problem  can be mapped on
a single channel Kondo model when the dissipation is of
ohmic type\cite{sudip}.  The hamiltonian reads
\begin{eqnarray}
H_\lambda&=&\int_{-\infty}^0 dx \frac{1}{2}[\Pi^2+(\partial_x\phi)^2]
+
\nonumber \\
& & \lambda \delta(x) (S_+ e^{i\sqrt{2\pi g}\phi}+
S_- e^{-\sqrt{2\pi g}\phi}).
\end{eqnarray}
The spins $S_\pm$ are spin $1/2$ operators,
and the $S_z$ value corresponds to the two states of the
dissipative model.  The dimensionless number $g$ characterizes
the dissipation.  The quantity $P(t)$ was defined in the
introduction;
it reads
\begin{equation}
P(t)={}\langle 0|S_z(t) |0\rangle
\label{pdef}
\end{equation}
where $|0\rangle$ is the product $|0\rangle_{\lambda=0}\otimes
|+\rangle$,
$|0\rangle_{\lambda=0}$  the ground state of the theory with
$\lambda=0$.
The Heisenberg operator $S_z(t)$ in (\ref{pdef})
evolves with  $H_\lambda$. To compute $P(t)$, we
 insert a complete set
of eigenstates of $H_\lambda$ on each side of $S_z$. These
are multiparticle states resulting from the massless limit of the
 sine-Gordon model \cite{fendley}; the spectrum is richer
than in the Ising case and consists
of solitons/anti-solitons, and, for $g<1/2$, of breathers
which we denote by $\epsilon=\pm, 1,2,...$ respectively.
The determination of $P(t)$ then
requires the  scalar products
${}^{\epsilon_p,...,\epsilon_1}_\lambda\langle\beta_p,...,
\beta_1|0\rangle$, which are of the same nature as those discussed
previously, together with the form factors of the
spin operator in the massless limit.

A subtle point has to be emphasized here:  the integrable picture is
an infrared
one, where the spin is  screened; that is, it doesn't appear in
the multiparticle description.
As demonstrated in \cite{skorik} its properties  can nevertheless be
computed by using correlators
of the current operator $\partial_x\phi$ as follows
\begin{equation}
\label{relat}
S_z(t)-S_z(0)=\int_0^t dt' \partial_x\phi(x=0,t').
\end{equation}
That the spin is  up in the initial state is then taken into account
by giving a unit charge
to the state $|0\rangle$; the non vanishing scalar products
are thus those for which $\sum \epsilon_i=1$.

At $g=1/2$ the sine-Gordon model decouples into two
massless Ising models with a boundary magnetic field
$h\propto \lambda$ and the previous considerations
on the Ising model can be used.  There are
no bound states at that point and the
reflection matrix for the solitons and anti-solitons
is
\begin{equation}
R_{+-}=R_{-+}=i\frac{e^\beta-iT_b}{e^\beta+iT_b}, \
R_{++}=R_{--}=0.
\label{rmat}
\end{equation}
$T_b$  is describing the boundary scale and is related
to $\lambda$, the precise relation to be found in
Eq. (5.7) of \cite{voltage}.  The current operator
has a simple form factor at that point given by (we turn to real time
here)
\begin{eqnarray}
{}_\lambda\langle 0|\partial_x\phi&|&\beta_1,\beta_2
\rangle_\lambda^{\epsilon_1,\epsilon_2}\propto
\delta_{\epsilon_1+\epsilon_2} \epsilon_1 e^{(\beta_1+\beta_2)/2}
\\
& & [1+R_{+-}(\beta_1) R_{+-}(\beta_2)] e^{-it(e^{\beta_1}+
e^{\beta_2})}.
\end{eqnarray}
We won't need the detailed knowledge of the scalar products
(\ref{scalar});
only their properties under crossing, together with  the value of the
residues at the kinematical poles, matter.

Many processes contribute to the evaluation of $P(t)$.
The simplest takes the form
\begin{equation}
\int \frac{d\beta_1d\beta_2}{(2\pi)^2} \
{}_0\langle 0|\beta_2\rangle^+_\lambda {}_\lambda^+\langle
\beta_1|0\rangle_0 \ {}_\lambda^+\langle\beta_2|
\partial_x\phi(t)|\beta_1\rangle_\lambda^+.
\end{equation}
This process gets convoluted with many others,
that is one has to add to the factor
$\langle 0|\beta_2\rangle_\lambda^+\
{}_\lambda^+\langle\beta_1|0\rangle$
the sum
\begin{equation}
\int {d\beta_3d\beta_4\over (2\pi)^2}
\langle 0|\beta_4\beta_3\beta_2\rangle_\lambda^{+-+}\ \
{}_\lambda^{+-+}\langle\beta_4\beta_3\beta_1|0\rangle
+\ldots.
\end{equation}
Suppose we move the rapidity integrals to $\hbox{Im }\beta_1=-i\pi$
and $\hbox{Im }\beta_2=i\pi$. Forgetting the singularities
encountered in doing so, one obtains, using crossing, the scalar
product
${}_\lambda^+\langle\beta_2|\beta_1
\rangle_\lambda^+=2\pi\delta(\beta_1-\beta_2)$.
But one easily cheks that the form factor
${}_\lambda^+\langle\beta|\partial_x\phi|\beta\rangle_\lambda^+$
vanishes, so the whole series adds up to sero! All
what matters therefore are the contributions of the poles encountered
in
moving the contours.
Moreover, one can show that the kinematical poles do not contribute;
the one point function
is entirely determined by the poles of the
the  reflection matrix  at $\beta_{1,2}=\log(T_b)\mp i\pi/2$.
and one gets the well known result \cite{sudip}
\begin{equation}
P(t)=e^{-2 T_b t}.
\end{equation}

The main feature of this computation - the fact that $P(t)$ is
entirely
determined by the poles of the $R$ matrices - generalizes to
arbitrary values of $g$, since it follows entirely from the
g-independent
general properties of the scalar products. As a result, we can
immmediately
obtain interesting properties. For $g>1/2$ there are  only solitons
and anti-solitons
in the spectrum, for which  the reflection matrix is
independent of $g$, and still given by (\ref{rmat}).
It follows that the behaviour is entirely incoherent,
and that $P(t)$ has the form given in (\ref{poft}). All terms now
contribute
since the form factor of $\partial_x\phi$ is non zero for
 any even number of particles
at $g>{1\over 2}$.

When $g<1/2$ there
are also $m < [1/g]$ breathers in the spectrum
(m integer),with  a  reflection matrix
given by\cite{gho}
\begin{equation}
R_m(\theta)=-{\tanh(\frac{\beta-\beta_b}{2}-i\frac{
\pi m g}{4(1-g)})\over
\tanh(\frac{\beta-\beta_b}{2}+i\frac{\pi m g}{4(1-g)})},
\end{equation}
with $\beta_b=\log T_b$.
This matrix has poles where $e^\beta$ has  non vanishing
real and imaginary part: they give rise to oscillatory
contributions to $P(t)$, indicating that the behaviour is coherent
in that domain.  An expression similar to (\ref{poft}) can be
written:
the leading behaviour at large times (or small $g$) follows from the
one breather,
as given in (\ref{poftasymp}), leading to the ratio (\ref{ratio}).
This result is consistent with the expansion in
$g=1/2-\epsilon$ done in \cite{Egger}. It also agrees with the $g\to
0$ limit
in which $\lambda\approx \pi g T_b/2$ \cite{voltage}.
It would be very interesting to test this numerically. Formula
(\ref{poftasymp})
should be especially useful in the quantum optics context, where the
values of $g$ are usually quite small.


\begin{references}

\bibitem{ian} I. Affleck, A. W. W. Ludwig, Nucl. Phys. B360 (1991)
641; Nucl. PHys. B428 (1994)
545.

\bibitem{bethe} P. B. Wiegmann, A. M. Tsvelick, JETP Lett. 38 (1983)
591; N. Andrei,
K. Furuya, J. Lowenstein, Rev. Mod. Phys. 55 (1983) 331.

\bibitem{fendley} P. Fendley, A.W.W. Ludwig, H. Saleur, Phys.
Rev. Lett. 75 (1995) 2196.

\bibitem{fendleyi} P. Fendley, A. W. W. Ludwig, H. Saleur, Phys. Rev.
B52 (1995) 8934.

\bibitem{korepin} V.E. Korepin, N.M. Bogoliubov, A.G. Izergin,
``Quantum
inverse scattering method and correlation functions:,
(Cambridge university press), Cambridge 1993.

\bibitem{exps} F.P. Milliken, C.P. Umbach, R.A. Webb, Solid State
Comm. {\bf 97}, (1996) 309.

\bibitem{expsi} L. Saminadayar, D.C. Glattli, Y. Jin, B. Etienne,
cond-mat/9706307; to appear Phys. Rev. Lett.

\bibitem{Barry} B. Mc Coy, T.T. Wu, ``The two dimensional Ising  
model'',
Oxford University Press (1973)

\bibitem{ghozamo} S. Ghoshal, A.B. Zamolochikov, Int. J. Mod. Phys. A
{\bf 9}, (1994) 3841.

\bibitem{smirnov} F.A. Smirnov, ``Form factors in completely
integrable models of quantum field theory", World scientific
(Singapore) and references therein.

\bibitem{giuseppe} G. Delphino, G. Mussardo, P. Simonetti, Phys.
Rev. {\bf D}51, (1995) 6620.

\bibitem{skorik} F. Lesage, H. Saleur, S. Skorik, Phys. Rev. Lett.
76, (1996) 3388, cond-mat/9512087; Nucl. Phys. B 474 [FS],
(1996) 602, cond-mat/9603043.

\bibitem{kosti} T. A. Costi and C. Kieffer, Phys. Rev. Lett. 76
(1996) 1683.

\bibitem{sudip} A. J. Leggett, S. Chakravary, A.T. Dorsey, M. P. A.
Fisher, A. Garg, and W. Zwerger, Rev. Mod. Phys. 59 (1987) 1.


\bibitem{sudipi} S. Chakravarty, J. Rudnick, Phys. Rev. Lett.
{\bf 75}, (1995) 701.

\bibitem{eggr} R. Egger, H. Grabert, U. Weiss, Phys. Rev.
E {\bf 55},  (1997) 3809.

\bibitem{Egger} R. Egger, H. Grabert, Phys. Rev. B55 (1997) R3809.

\bibitem{Strong} S.P. Strong, cond-mat/9702141.

\bibitem{seethna} A.A.  Louis, J.P.  Seethna, Phys. Rev. Lett. 74,
1363 (1995)

\bibitem{leclair} A. LeClair, F. Lesage, S. Lukyanov, H. Saleur,
Phys. Lett. A235, 1997) 203; hep-th/970122.

\bibitem{JCardy} J. Cardy, Nucl. Phys. {\bf B}324, (1989) 581.

\bibitem{giuseppei} R. Konik, A. LeClair, G. Mussardo,
Int. J. Mod. Phys. A11, (1996) 2765.

\bibitem{nous}  F. Lesage, H. Saleur, To appear.


\bibitem{Burk} T. W. Burkhardt, T. Xue, Nucl. Phys. B354 (1991) 653.

\bibitem{gho} S. Ghoshal, Int. J. Mod. Phys. A{\bf 9}, (1994) 4801.

\bibitem{voltage}  F. Lesage, H. Saleur, Nucl. Phys. B490  [FS]
(1997)  543.






\end{references}
\end{document}